\documentclass[10pt,pre,twocolumn,showpacs,aps,floats,floatfix,superscriptaddress]{revtex4}

\usepackage{epsfig}
\begin{document}

\newcommand{\avk}{\langle k \rangle}
\newcommand{\fluck}{\langle k^2 \rangle}

\title{The effects of spatial constraints on the evolution of weighted complex networks}

\author{Alain Barrat} 
\affiliation{Laboratoire de Physique Th\'eorique (UMR du CNRS 8627),
B\^atiment 210, Universit{\'e} de Paris-Sud 91405 Orsay, France}
\author{Marc Barth\'elemy\footnote{On leave of absence from 
CEA-Centre d'Etudes de Bruy{\`e}res-le-Ch{\^a}tel, 
D\'epartement de Physique Th\'eorique et
Appliqu\'ee BP12, 91680 Bruy\`eres-Le-Ch\^atel, France}}
\affiliation{School of Informatics and Biocomplexity Center,
Indiana University,Eigenmann Hall, 1900 East Tenth Street,
Bloomington, IN 47406}
\author{Alessandro Vespignani}
\affiliation{School of Informatics and Biocomplexity Center, 
Indiana University,Eigenmann Hall, 1900 East Tenth Street, 
Bloomington, IN 47406} 
\date{\today} \widetext
\begin{abstract}

Motivated by the empirical analysis of the air transportation system,
we define a network model that includes geographical attributes along
with topological and weight (traffic) properties.  The
introduction of geographical attributes is made by constraining
the network in real space. Interestingly, the inclusion of geometrical
features induces non-trivial correlations between the weights, the
connectivity pattern and the actual spatial distances of vertices.
The model also recovers the emergence of anomalous fluctuations in the
betweenness-degree correlation function as first observed by Guimer\`a
and Amaral [Eur. Phys. J. B {\bf 38}, 381 (2004)].  The presented
results suggest that the interplay between weight dynamics and spatial
constraints is a key ingredient in order to understand the formation
of real-world weighted networks.

\end{abstract}


\maketitle 

\section{Introduction}

The empirical evidence coming from studies on systems belonging to areas as
diverse as social sciences, biology and computer science have shown that
the usual paradigm of random graphs is often not well suited to
describe real world networks~\cite{barabasi:2002,dorogovtsev:2002,mdbook,psvbook}. 
In particular, in a wide range of networks the occurrence of vertices 
with a very large degree (number of links to other vertices) is very likely.
The presence of these ``hubs'' often goes along with very large 
degree fluctuations. 
The large topological heterogeneity associated to these features 
is statistically expressed by the presence of heavy-tailed degree 
distributions with diverging variance that have a very strong impact on
the networks' physical properties such as resilience and
vulnerability, or the propagation of pathogen agents
\cite{havlin00,newman00,barabasi00,pv01a}.

The purely topological definition of networks, however, misses 
important attributes which are frequently encountered in real-world
networks. In the first instance, networks are far from boolean
structure and are better represented as weighted graphs 
with the intensity of links that may vary over many orders of
magnitude. Indeed, in many graphs ranging from food-webs to metabolic
networks, large variations of
the link intensities are empirically observed 
~\cite{granovetter:1973,garla:2003,li:2003a,li:2003b,krause:2003,barrat:2004a,almaas:2004}.
Notably, the statistical properties of weights indicate non-trivial correlations
and association with topological quantities~\cite{barrat:2004a}. Finally, the
correlation between weights on different links is at the origin of the
existence of pathways which are particularly important in metabolic
networks for example~\cite{almaas:2004}.  
Another important element of many real networks is their embedding
in the real space. For instance, most people have their friends and
relatives in their neighborhood, transportation networks depend on
distance, and many communication networks devices have short radio
range~\cite{helmy02,nemeth02,gorman03,Gastner:2004a,Gastner:2004b}. A
particularly important example of such a ``spatial'' network is the
Internet which is a set or routers linked by physical cables with
different lengths and latency times~\cite{lakhina02,psvbook}. An
analogous situation is faced in the air transportation network with
routes covering very different distances. The length of the link is a
very important quantity usually associated with an intrinsic cost in
the establishment of the connection. If the cost of a
long-range link is high, most of the connections starting from a given
node will go to the closest neighbors in the embedding space.
Long-range links, on the other hand, correspond usually to connections
towards already well-connected nodes (hubs).  This seems natural in
the case of the air transportation network for instance: short
connections go to small airports while long distance flights are
directed preferentially towards large airports ({\it i.e.} well
connected nodes). It is therefore natural to find that spatial 
constraints can have important consequences on the topology of the 
resulting network~\cite{barthelemy:2003a,amaral:2004}.

Recently, the raising interest on the dynamics and function of complex
networks has fostered studies going beyond the simple topological
structure. In particular, models of complex 
networks in which the diversity of weights is taken into account have
been formulated having in mind growing networks where the dynamics is 
driven by the intensity of the weights along with a reinforcement mechanism  
~\cite{barrat:2004b,barrat:2004c,barrat:2004d,pandya:2004}. 
Other models have focused on more geometrical mechanisms or somehow different
dynamical rules~\cite{dorogovtsev:2004,krapivsky:2004,bianconi:2004,wang:2005}.
These models, however, are not able to reproduce all of the features
observed in real world networks. For instance, the
anomalous centrality fluctuations observed in Ref.~\cite{amaral:2004}
do not find a rationalization in models based only on topology and
weight properties. On the other hand, some of
these interesting and non-trivial features can result from
the introduction of spatial attributes in the models' construction
\cite{amaral:2004}. 
In this article, we discuss the interplay of the 
three aforementioned ingredients (heterogeneous topology, weights and spatial
constraints) in a model of growing network combining these ingredients
at once. The proposed model is obtained as the
embedding of the weighted growing network introduced
in~\cite{barrat:2004b} in a 
two-dimensional geometrical space. Spatial constraints
are translated into a preference for short links, and combined with
the coupling between the evolution of the network and the dynamical
rearrangement of the weights. This mechanism naturally leads to the
appearance of many features observed in real-world networks, in
particular the non-linear correlations between weights and topology,
and the large fluctuations of the betweenness centrality.

The paper is organized as follows. In section \ref{sec:II}, we briefly
review some important empirical results of the North-American airline
network, highlighting the most salient effects on space on different
quantities. Sections \ref{sec:III} and \ref{sec:IV} are devoted to the
presentation and to the study of the spatial weighted model, stressing
the effect of the spatial embedding and constraints on the properties
of the resulting network. In section \ref{sec:V}, we present a
summary of the results and conclusions about large network modeling.

\section{A case study: Space, topology and traffic in the North
American airline network}
\label{sec:II}

The airline transportation infrastructure is a paramount example of large
scale network  which can be represented as a complex weighted graph:
the airports are the vertices of the graph and the links represent the
presence of direct flight connections among them. 
The weight on each link is the
number of maximum passengers on the corresponding connection.  The
characteristics of the world-wide air-transportation network using the
International Air Transportation Association (IATA) database
\cite{IATA} have been presented in~\cite{barrat:2004a}. The network
is made  of $N=3880$ vertices and $E=18810$ edges and shows both small-world and
scale-free properties as also confirmed in different datasets and analyses
\cite{li:2003a,li:2003b,guimera:2003,amaral:2004}. In particular, the
average shortest path length, measured as the average number of edges
separating any two nodes in the network shows the value
$\langle\ell\rangle=4.37$, very small compared to the network size
$N$. The degree distribution,  takes the form $P(k)=
k^{-\gamma}f(k/k_x)$, where $\gamma\simeq2.0$ and $f(k/k_x)$ is an
exponential cut-off function. The degree distribution is therefore
heavy-tailed with a cut-off that finds its origin in the physical
constraints on the maximum number of connections that a single airport
can handle~\cite{guimera:2003,amaral:2004,amaral:2000}. The airport connection
graph is therefore a clear example of small-world network showing
a heavy-tailed degree distribution and heterogeneous topological 
properties. 

The world-wide airline network necessarily mixes different effects. In
particular there are clearly two different scales, global
(intercontinental) and domestic. 
The intercontinental scale defines two different groups of travel 
distances and for the statistical consistency we eliminate this 
specific geographical constraint by focusing on a single 
continental case. Namely, in the following we will consider 
the North-American network constituted of $N=935$ vertices with an average
degree $\langle k\rangle \approx 8.4$ and an average shortest path
$\ell\approx 4$. The statistical topological properties of the North American
network are consistent with the world-wide one as we will see in the
forthcoming analysis.

\subsection{Topology and weights}

\begin{figure}[t]
\vskip .5cm
\begin{center}
\epsfig{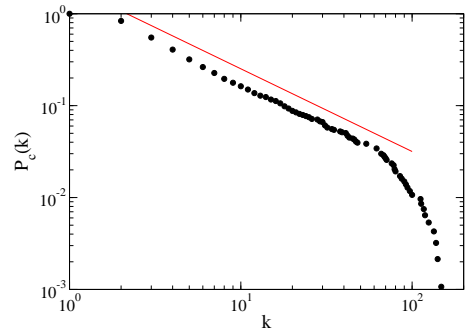}
\end{center}
\caption{Cumulative degree distribution $P_c(k)$ for the
North-American network. The straight line indicates a power-law decay
with exponent $\gamma - 1=0.9$.}
\label{figNAdegree}
\end{figure}

The North American network presents a degree distribution
statistically consistent with the world-wide airline network.  Indeed,
we also observe (Fig.~\ref{figNAdegree}) in this case a power-law
behavior on almost two orders of magnitude followed by a cut-off indicating the maximum
number of connections possible due to limited airport capacity and to the size of
the network considered.

The existence of a broad degree distribution signals
a strong heterogeneity of the network at the topological level
which exists also at the weight level. 
A first indication on the weight heterogeneity is given by the study of
the weight strength of a node $i$ as defined by~\cite{yook:2001,barrat:2004a}
\begin{equation}
s_i^w=\sum_{j \in {\cal V}(i)} w_{ij} \ .
\end{equation}
where the sum runs over the set ${\cal V}(i)$ of neighbors of $i$. The
strength generalizes the degree to weighted networks and in the case
of the air transportation network quantifies the traffic of passengers
handled by any given airport. This quantity obviously depends on the
degree $k$ and increases (linearly) with $k$ in the case of random
uncorrelated weights of average $\langle w\rangle$. A relation between
the average strength $s^w(k)$ of nodes of degree $k$ of the form
\begin{equation}
s^w = A k^{\beta_w} \ ,
\label{eq:sw}
\end{equation}
with an exponent $\beta_w >1$, or $\beta_w=1$ but $A \ne \langle w
\rangle$ is then the signature of non-trivial statistical correlations between weights
and topology.  This is indeed what we observe in the North-American
air transportation network with $\beta_w \simeq 1.7$
(Fig.~\ref{sw-sd}).

\subsection{Spatial analysis}

\begin{figure}[t]
\vskip .5cm
\begin{center}
\epsfig{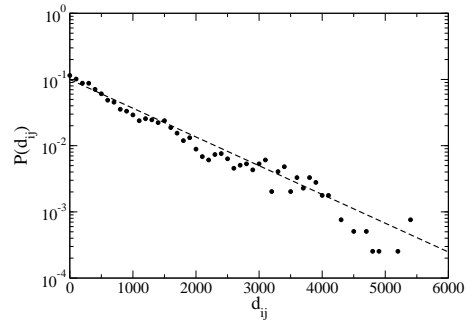}
\end{center}
\caption{Distribution of distances (in kms) between airports
linked by a direct connection for the North-American network. The
straight line indicates an exponential decay with scale of order
$1000$ km.}
\label{fig:Pd}
\end{figure}

The spatial attributes of the North American airport network are 
embodied in the physical spatial distance, measured in kilometers or
miles, characterizing each connection. Fig.~\ref{fig:Pd} displays the 
histogram of the distances of the direct flights. These distances correspond to
Euclidean measures of the links and clearly show a fast decaying
behavior reasonably fitted by an exponential. The exponential fit
gives a value for a typical scale of the order $1000$ kms. The origin
of the finite scale can be traced back to the existence of physical
and economical restrictions on airline planning in a continental setting.

\begin{figure}[t]
\vskip .5cm
\begin{center}
\epsfig{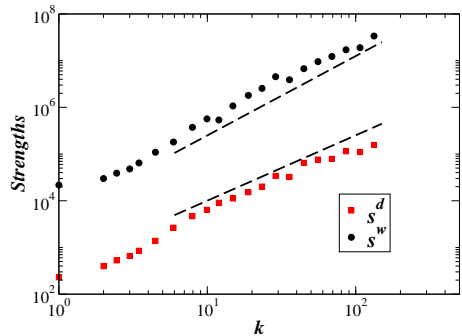}
\end{center}
\caption{ Weight and distance strengths versus degree 
  for the North-American network. The dashed lines correspond to the
  power-laws $\beta_d\simeq 1.4$ and $\beta_w\simeq 1.7$.}
\label{sw-sd}
\end{figure}

Since space is an important parameter in this network, another interesting
quantity is the {\em distance strength} of $i$
\begin{equation}
s_i^d=\sum_{j \in {\cal V}(i)} d_{ij}
\end{equation}
where $d_{ij}$ is the {\em Euclidean} distance.  This quantity gives
the cumulated distances of all the connections from (or to) the
considered airport.  Similarly to the usual weight strength,
uncorrelated random connections would lead to a linear behavior of
$s^d(k)\propto k$ while we observe in the North-American network a
power law behavior
\begin{equation}
s^{d}(k) \sim k^{\beta_d}
\label{eq:sd}
\end{equation}
with $\beta_d\simeq 1.4$ (Fig.~\ref{sw-sd}). This result shows
the presence of important correlations between topology and geography.
Indeed, the fact that the exponents appearing in the relations
(\ref{eq:sw}) and (\ref{eq:sd}) are larger than one have different
meanings. While Eq.~(\ref{eq:sw}) means that larger airports have
connections with larger traffic, (\ref{eq:sd}) implies that they have
also farther-reaching connections. In other terms, the traffic (and the distance) per
connection is not constant but increases with $k$.  As intuitively expected, the
airline network is an example of a very heterogeneous network where
the hubs have at the same time large connectivities, large weight
(traffic) and long-distance connections~\cite{barrat:2004a}, related
by super-linear scaling relations.

\subsection{Assortativity and Clustering}

A complete characterization of the network structure 
must take into account the level of degree correlations and clustering
present in the network. 
Correlations can be probed by 
inspecting the average degree of the nearest neighbors
of a vertex $i$ 
\begin{equation}
k_{nn,i}=\frac{1}{k_i}\sum_{j\in {\cal V}(i)} k_j \ .
\end{equation}
Averaging this quantity over nodes with same degree $k$ leads to a convenient measure 
to investigate the behavior of the degree correlation function~\cite{psvbook,vazquez02} 
\begin{equation}
k_{nn}(k) = \frac{1}{N_k} \sum_{i/k_i=k} k_{nn,i}\ ,
\label{knnw}
\end{equation}
where $N_k$ is the number of nodes of degree $k$. This quantity (\ref{knnw}) is
related to the correlations between the degrees of connected vertices
since on average it can be expressed as
\begin{equation}
k_{nn}(k) = \sum_{k'} k' P(k'|k) \ .
\end{equation}
where $P(k'|k)$ is the conditional probability that a given vertex with
degree $k$ is linked to a vertex of degree $k'$.
If the degrees of neighboring vertices are uncorrelated,
$P(k'|k)$ is only a function of $k'$ and thus $k_{nn}(k)$ is a
constant. When correlations are present, two main classes of possible
correlations have been identified: {\em Assortative} behavior if
$k_{nn}(k)$ increases with $k$, which indicates that large degree
vertices are preferentially connected with other large degree
vertices; and {\em disassortative} if $k_{nn}(k)$ decreases with
$k$~\cite{Newman:2002}.  The weighted generalization of the above
quantity, the affinity, reads as~\cite{barrat:2004a}
\begin{equation}
  k^w_{nn,i}=\frac{1}{s_i}\sum_{j \in \mathcal{V}(i)} w_{ij} k_j.
\end{equation}
In this case, we perform a local weighted average of the nearest
neighbor degree according to the normalized weight of the connecting
edges, $w_{ij} / s_i$.  This definition implies that $k^w_{nn,i}>
k_{nn,i}$ if the edges with the larger weights are pointing to the
neighbors with larger degree and $k^w_{nn,i}< k_{nn,i}$ in the
opposite case. The $k^w_{nn,i}$ thus measures the effective {\em
affinity} to connect with high or low degree neighbors according to
the magnitude of the actual interactions.  As well, the behavior of
$k^w_{nn}(k)$, i.e.  the affinity of vertices of degree $k$,
marks the weighted assortative or
disassortative properties considering the actual interactions among
the system's elements.

Information on the local connectedness is provided by the clustering
coefficient $c_i$ defined for any vertex $i$ as the fraction 
of connected neighbors of
$i$~\cite{watts98}. The average clustering coefficient ${\cal C}=N^{-1}\sum_i
c_i$ thus expresses the statistical level of cohesiveness measuring
the global density of interconnected vertices' triples in the networks,
and the function $C(k)$ restricted to classes of vertices with degree $k$
allows to gather more detailed information.
A possible weighted definition of the clustering coefficient is
provided by the expression
\begin{equation}
  c^w(i)=\frac{1}{s_i(k_i-1)} \sum_{j, h \in {\cal V}(i)}
  \frac{(w_{ij}+w_{ih})}{2} a_{jh},
\end{equation}
where $a_{jh}$ is equal to $1$ if $j$ and $h$ are linked, and $0$
otherwise. The quantity $c^w(i)$ takes into account the weight of the two
participating edges of the vertex $i$ for each triple formed in the
neighborhood of the vertex $i$; it measures the relative weight of
the triangles in the neighborhood of a vertex $i$ with respect to the
vertex' strength~\cite{barrat:2004a}. ${\cal C}^w$ and $C^w(k)$ are defined
as the average over all nodes and over nodes of degree $k$, respectively.
It is worth remarking that
alternative definitions of the weighting scheme for clustering have
been proposed in the literature~\cite{onnela:2004}.

\begin{figure}[t]
\vskip .5cm
\begin{center}
\epsfig{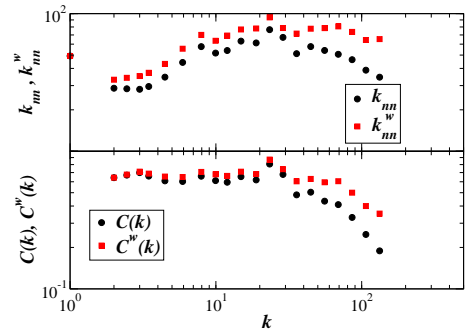}
\end{center}
\caption{Assortativity and clustering for the North-American network.
Circles correspond to topological quantities while squares are for
affinity and weighted clustering.
}
\label{fig:Assor}
\end{figure}

Figure~\ref{fig:Assor} displays for the North-American airport network
the behavior of these various quantities as a function of the
degree. An essentially flat $k_{nn}(k)$ is obtained and a slight
disassortative trend is observed at large $k$, due to the fact that
large airports have in fact many intercontinental connections to other
hubs which are located outside of North America and are not considered
in this ``regional'' network. The clustering is very large and
slightly decreasing at large $k$. This behavior is often observed in
complex networks and is here a direct consequence of the role of large
airports that provide non-stop connections to different regions which
are not interconnected among them. Moreover, weighted correlations are
systematically larger than the topological ones, signaling that large
weights are concentrated on links between large airports which form
well inter-connected cliques (see also~\cite{barrat:2004a} for more details).

\subsection{Betweenness Centrality}

\begin{figure}[t]
\vskip .5cm
\begin{center}
\epsfig{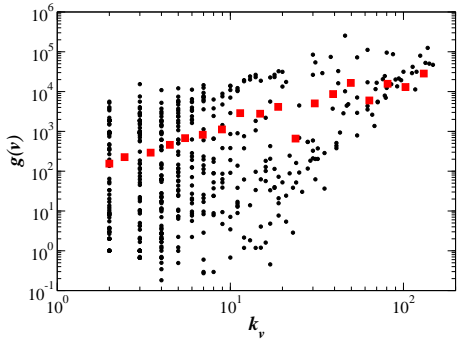}
\end{center}
\caption{Scatter-plot of the betweenness centrality versus degree
for nodes of the North-American air-transportation network. The red squares
correspond to the average BC versus degree.}
\label{fig:bc_na}
\end{figure}

A further characterization of the network is provided by considering
quantities that takes into account the global topology of the
network. For instance, the degree of a vertex is a local measure that
gives a first indication of its centrality. However, a more
global approach is needed in order to characterize the real importance
of various nodes. Indeed, some particular low-degree vertices may be
essential because they provide connections between otherwise separated
parts of the network. In order to take properly into account such
vertices, the betweenness centrality (BC) is commonly used
\cite{freeman77,newman01,goh01,barthelemy:2003b}.  The betweenness
centrality of a node $v$ is defined as
\begin{equation}
g(v)=\sum_{s\neq t}\frac{\sigma_{st}(v)}{\sigma_{st}}
\end{equation}
where $\sigma_{st}$ is the number of shortest paths going from $s$ to $t$ 
and $\sigma_{st}(v)$ is the number of shortest paths going from $s$ to $t$ 
and passing through $v$. This definition means that
central nodes are part of more
shortest paths within the network than peripheral nodes. Moreover, the
betweenness centrality gives in transport networks an estimate
of the traffic handled by the vertices, assuming that the
number of shortest paths is a zero-th order approximation to the
frequency of use of a given node.
It is generally useful to represent the average betweenness centrality 
for vertices of the same degree
\begin{equation}
g(k)=\frac{1}{N_k}\sum_{v/k_v=k}g(v) \ .
\end{equation}
For most
networks, $g(k)$ is strongly correlated with the degree $k$. In
general, the largest the degree and the largest the
centrality. For scale-free networks it has been shown that the
centrality scales with $k$ as
\begin{equation}
g(k) \sim k^\mu
\label{eq:gk}
\end{equation}
where $\mu$ depends on the network
\cite{newman01,goh01,barthelemy:2003b}.  
For some networks however, the BC fluctuations around the behavior given by
Eq.~(\ref{eq:gk}) can be very large and ``anomalies'' can occur, in
the sense that the variation of the centrality versus degree is
not a monotonous function.  Guimer\`a and
Amaral~\cite{amaral:2004} have shown that this is indeed the case 
for the air-transportation network. This is a very relevant observation
in that very  central cities may have a relatively low
degree and vice versa. In Fig.~\ref{fig:bc_na} we report
the average behavior along with the scattered plot of the betweenness
versus degree of all airports of the North American network. Also in
this case we find very large fluctuations with a behavior similar to
those observed in Ref.~\cite{amaral:2004}. Interestingly, Guimer\`a and
Amaral have put forward a network model embedded in real space
that considers  geopolitical constraints. This model appears to
reproduce the betweenness centrality features observed in the real
network pointing out the importance of space as a relevant ingredient
in the structure of networks. In the following we focus on the
interplay between spatial embedding, topology and weights in a simple
general model for weighted networks in order to provide a modeling
framework considering these three aspects at once.


\section{The model}
\label{sec:III}

Early modeling of weighted networks just considered weight and
topology as uncorrelated quantities~\cite{yook:2001}. This is not the
case in real world networks where a complex interplay between the
evolution of weights and topological growth does exist. For instance, if a new airline
connection between two airports is created, it will generally provoke
a modification of the existing traffic of both airports. In general,
it will increase the traffic activity depending on the specific nature
of the network and on the local dynamics. 
This effect is introduced in the modeling of growing networks by 
the mechanism of a strength preferential
attachment together with a dynamical redistribution of
weights. Following this strategy it is possible to 
produce weighted networks with broad distributions of weights,
connectivities, and strengths, and correlations between weights and
topology~\cite{barrat:2004b,barrat:2004c}.

Here we consider a weighted growing network whose nodes are embedded
in a two-dimensional space. As in the weighted model of
Ref.~\cite{barrat:2004b}, it is reasonable to think that a newly
created node $n$ will establish links towards pre-existing nodes with
heavy traffic or strength (hubs).  Costs are however associated with
distances and there is a trade-off between the need to reach a hub in
a few hops and the connection costs. The cost naturally increases with
the distance implying that the probability of establishing a
connection between the new node $n$ and a given vertex $i$ decays as a
function of the increasing Euclidean distance $d_{ni}$. As in the case
of topological preferential attachment (i.e. connecting probability
proportional to the degree~\cite{barabasi:1999}), this trade-off can
be expressed in two different ways: the connecting probability can
decrease either as a power-law of the
distance~\cite{manna02,xulvi02,yook02} or as an exponential with a
finite typical scale~\cite{barthelemy:2003a} as it seems more natural
for networks such as transportation networks (see Fig.~\ref{fig:Pd})
or technological networks~\cite{waxman88}.  All the effects described
in the next paragraphs are obtained in the case of an exponential
decay $\exp (-d_{ni}/r_c)$ but are also present in the case of a
power-law $d_{ni}^{-a}$ (the effect of a decreasing scale $r_c$ is
qualitatively the same as the effect of an increasing exponent
$a$). Eventually, the creation of new edges will introduce new traffic
which will trigger perturbations in the network. This model therefore
consists of two combined mechanisms:
\begin{enumerate}
\item{} {\it Growth}. We start with an initial seed of $N_0$ vertices
randomly located (with uniform distribution) on a $2$-dimensional disk
(of radius $L$) and connected by links with assigned weight $w_0$. At
each time step, a new vertex $n$ is placed on the disk at a randomly
assigned position ${\bf x}_n$ (still according to a uniform
distribution). This new site is connected to $m$ previously existing
vertices, choosing preferentially nearest sites with the largest
strength.  More precisely, a node $i$ is chosen according to the
probability
\begin{equation}
\Pi_{n \to i}=\frac{s_i^w e^{-d_{ni}/r_c}}{\sum_j s_j^w e^{-d_{nj}/r_c}} ,
\label{sdrive}
\end{equation}
where $r_c$ is a typical scale and $d_{ni}$ is the Euclidean distance
between $n$ and $i$. This rule of {\em strength driven
preferential attachment with spatial selection}, generalizes the
preferential attachment mechanism driven by the strength to spatial
networks. Here, new vertices connect more likely to vertices which
correspond to the best interplay between Euclidean distance and strength.
\item{} {\it Weights dynamics}. 
The weight of each new edge $(n,i)$ is fixed to a given value $w_0$
(this value sets a scale so we can take $w_0=1$). The creation of this
edge will perturb the existing interactions and we consider {\em
local} perturbations for which only the weights between $i$ and its
neighbors $j\in{\cal V}(i)$ are modified
\begin{equation}
w_{ij}\to w_{ij}+\delta\frac{w_{ij}}{s^{w}_i}.
\label{eq:rule}
\end{equation}
\end{enumerate}
After the weights have been updated, the growth process is iterated by
introducing a new vertex, i.e. going back to step (1.) until the
desired size of the network is reached.

The previous rules have simple physical and realistic interpretations.
Equation~(\ref{sdrive}) corresponds to the fact that new sites try to
connect to existing vertices with the largest strength, with the
constraint that the connection cannot be too costly. This adaptation
of the rule ``busy get busier'' introduced in
\cite{barrat:2004b,barrat:2004c} allows to take into account physical
constraints.  The weights' dynamics Eq.~(\ref{eq:rule}) expresses the
perturbation created by the addition of the new node and link. It
yields a global increase of $w_0 +\delta$ for the strength of $i$,
which will therefore become even more attractive for future nodes.

The value of $\delta$ characterizes the susceptibility of the
network.  If $\delta <w_0$, the new link does not have a large
influence. The case
$\delta \approx w_0$ corresponds to situations for which the new
created traffic (on the new link $n-i$) is transferred onto the
already existing connections in a ``conservative'' way. Finally,
$\delta >w_0$ is an extreme case in which a new edge generates a
sort of multiplicative effect that is bursting the weight or traffic
on neighbors.

The model contains two relevant parameters: the ratio between the typical
scale and the size of the system $\eta=r_c/L$, and the ability to redistribute
weights, $\delta$. Depending on the value of $\eta$ and $\delta$ we obtain different
networks whose limiting cases are summarized in Figure~\ref{fig:etadelta}.
More precisely, we expect:
\begin{figure}
\vspace*{.1cm} 
\centerline{
\epsfig{file=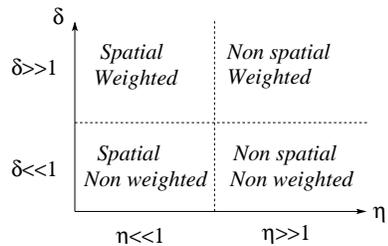,width=5cm}}
\caption{ Different limiting regimes of the model depending on the value
of its parameters $\delta$ and $\eta$.
}
\label{fig:etadelta}
\end{figure}
\begin{itemize}
  
\item{} For $\eta\gg 1$, the effect of distance is negligible and we
  recover the properties of the weighted model of
  Ref.~\cite{barrat:2004b}.  In this case, one obtains power-law
  distributions for connectivities and strengths with exponent
  $\gamma=(4\delta+3)/(2\delta+1)$, as well as for the weights
  (exponent $\alpha=2+1/\delta$). The strength and degree are linearly
  related by $s^w(k) \simeq (2\delta +1) k$.  The effect of the
  redistribution parameter $\delta$ is to broaden the various
  probability distributions, and to increase the correlations between
  topology and weights. Moreover, no correlations are introduced
  between the topology of the network and the underlying
  two-dimensional space, so that the distance strength $s^d$ grows
  simply linearly with the degree.

\item{} When $\eta$ decreases, additional constraints appear and have
  consequences that we will investigate numerically in the following. Unless
  otherwise specified, the simulations correspond to the parameters $m=3$
  (i.e. an average degree $\langle k \rangle = 6$), and $\delta=1.0$. We
  consider networks of size up to $N=10,000$, and the results are averaged
  over up to $100$ realizations. All the observed dependences in $\eta$ are
  essentially the same for other investigated values of $\delta$.

\end{itemize}

\section{Numerical results}
\label{sec:IV}

\subsection{Topology and weights}

\begin{figure}
\vspace*{.1cm} 
\centerline{
\epsfig{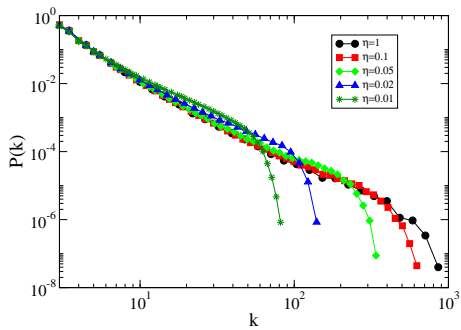}}
\vspace*{.5cm} 
\caption{Degree distribution $P(k)$
for different values of $\eta$ and $\delta=1$.
The degree distribution is averaged over $50$ networks of size $N=10^4$
and minimum degree $m=3$.
}
\label{fig:cutoff}
\end{figure}

\begin{figure}
\vspace*{.3cm} 
\centerline{
\epsfig{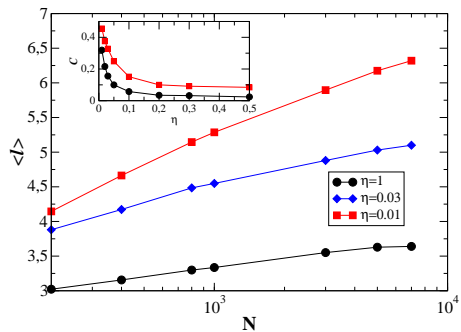}}
\vspace*{.5cm} 
\caption{Average shortest path distances as a function of the network size $N$
for different values of $\eta$, with $m=3$ and $\delta=1$.  Inset:
clustering coefficient as a function of $\eta$ for $N=10^4$,
$\delta=1$ (circles) and $\delta=5$ (squares). The data are averaged
over $50$ networks.  }
\label{fig:smallworld}
\end{figure}

At a purely topological level, the principal effect of a typical
finite scale $r_c$ in the creation of new connections is to introduce
a cut-off in the scale-free degree
distributions~\cite{barthelemy:2003a}. In Fig.~\ref{fig:cutoff} we
report the degree distribution for a fixed value of $\delta$ and
decreasing values $\eta$. A more pronounced cut-off appears at
decreasing value of $\eta$ signalling the onset of a trade-off between
the number of connections and their cost in terms of Euclidean
distance.  The small-world properties of the network are as-well
modified~\cite{barthelemy:2003a}: on the one hand, the increasing
tendency to establish connections in the geographical neighborhood
favors the formation of cliques and leads to an increase in the
clustering coefficient (see inset of Fig.~\ref{fig:smallworld}).  On
the other hand, this same tendency leads to an increase in the
diameter of the graph, measured as the average shortest path distances
between pairs of nodes. The diameter however still increases
logarithmically with the size of the graph, as shown in
Fig.~\ref{fig:smallworld}: the constructed networks do display the
small-world property, even if strong geographical constraints are
present.

\begin{figure}[t]
\begin{center}
\epsfig{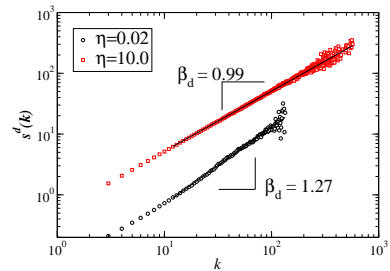}
\end{center}
\caption{ Distance strength versus $k$ for $\eta=0.02$ and 
$\eta=10.0$ (the networks are obtained for $\delta=1$, $N=10^4$,
  $\langle k\rangle =6$, and averaged over $100$ configurations). When
  $\eta$ is not small, space is irrelevant and there are no
  correlations between degree and space. When spatial effects
  are important ($\eta=0.02\ll 1$), non-linear correlations
  appear. We observe a crossover for $k\simeq 10-20$ to
  a power-law behavior and the power-law fit over this range of values of $k$ 
is shown (full lines).}
\label{fig:s.vs.k}
\end{figure}

The correlations appearing between traffic and topology of the network
are presented in Fig.~\ref{fig:s.vs.k} for two extreme cases of large
and small $\eta$. Strikingly, the effect of the spatial constraint is
to increase both exponents $\beta_w$ and $\beta_d$ to values larger
than $1$ and although the redistribution of the weights
[Eq.~(\ref{eq:rule})] is linear, non-linear relations
$s^w(k)$ and $s^d(k)$ as a function of $k$ appear. For the weight 
strength the effect is not very pronounced with an exponent
of order $\beta_w\approx 1.1$ for $\eta=0.01$, while for the distance
strength the non-linearity has an exponent of
order $\beta_d\approx 1.27$ for $\eta=0.02$ (the value of the exponents
$\beta_w$ and $\beta_d$ depend on $\eta$; see also
\cite{manna:2005} for a spatial model with 
$\beta_d > 1$). We show on Fig.~\ref{fig:s.vs.k}
the distance strength for two extreme situations for which spatial constraints
are inexistent ($\eta=10.0$) or on the contrary very strong ($\eta=0.02$).

The  nonlinearity induced by the spatial structure 
can be explained by the following mechanism affecting the network growth. 
The increase of spatial constraints affects the trend to
form global hubs, since long distance connections are less probable, 
and drives the topology towards the existence of
``regional'' hubs of smaller degree.  The total traffic however
is not changed with respect to the case $\eta=\infty$, and is in fact
directed towards these ``regional'' hubs. These medium-large degree
vertices therefore carry a much larger traffic than they would do if
global ``hubs'' were available, leading to a faster increase of the
traffic as a function of the degree, eventually resulting in a
super-linear behavior. Moreover, as previously mentioned,
the increase in distance costs implies that long range connections can
be established only towards the hubs of the system: 
this effect naturally lead to a super-linear accumulation of $s^d(k)$
at larger degree values.

\begin{figure}[t]
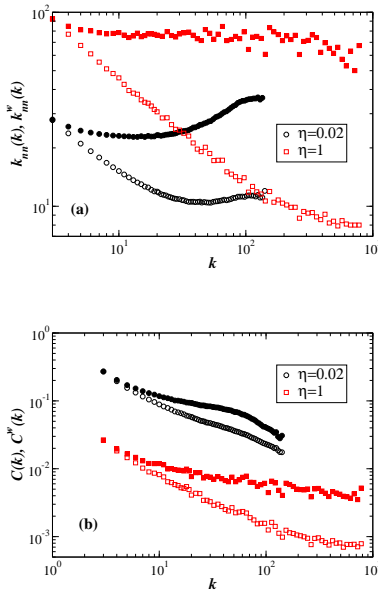

\begin{center}
\epsfig{file=./fig_knnknnw_model_delta1.eps,width=5cm}
\end{center}
\vskip .2cm
\begin{center}
\epsfig{file=./fig_ckckw_model_delta1.eps,width=5cm}
\end{center}
\caption{ (a) Assortativity and (b) clustering versus $k$ 
  obtained for the model for $\delta=1$, $N=10^4$, $\langle k\rangle =6$,
  $\eta=0.02$ (circles) and $\eta=1$ (squares). Data are averaged over $100$
  configurations. Empty symbols refer to topological correlations while full
  symbols correspond to the weighted quantities $k_{nn}^w$ and $C^w$.}
\label{fig:assor_mod}
\end{figure}

Spatial constraints have also a strong effect on the correlations
between neighboring nodes (Fig.~\ref{fig:assor_mod}). At large
$\eta$, a disassortative network is created, as is the case in most
growing networks~\cite{barrat:2004e}; as $\eta$ decreases, $k_{nn}$
decreases, and an increasing range of flat $k_{nn}(k)$ appears: the
tendency for small nodes to connect to hubs is contrasted by the need
to use small-range links.  For small enough $\eta$, a nearly neutral
behavior more similar to what is actually observed in the airport
network is reached. Moreover, the affinity of nodes to establish
strong links to large nodes, measured by $k_{nn}^w(k)$, goes from a
flat behavior at large $\eta$ to a slightly assortative one at small
$\eta$. In all cases, the weighted correlation $k_{nn}^w(k)$ remains
clearly larger than the unweighted $k_{nn}(k)$, showing that links to
busier nodes are typically stronger.

A non-trivial clustering hierarchy is already displayed by the model
without spatial constraints. As previously mentioned, the decrease of
$\eta$ leads to an increase of clustering. Moreover, the weighted
clustering is always significantly larger than the unweighted one,
showing that the cliques carry typically an important traffic
(see Fig.~\ref{fig:assor_mod}). These
effects are a general signature of spatial constraints as also
observed in a non weighted network~\cite{barthelemy:2003a}.

\subsection{Spatial constraints and betweenness centrality}

\begin{figure}
\vspace*{.1cm} 
\centerline{
\epsfig{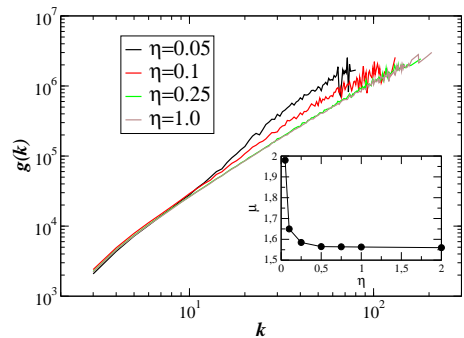}}
\vspace*{.5cm} 
\caption{ Betweenness centrality versus degree for 
different values of $\eta$. 
Inset: Exponent $\mu$ (obtained by fitting the data
for $k>10$) of the betweenness
centrality versus $\eta$ (for $N=5,000$, $m=3$ and averaged over $50$
configurations). For strong spatial constraints long-range shortcuts
are very rare and hubs connect regions which are otherwise almost
disconnected which in turn implies a larger centrality of the hubs.
}
\label{fig:muvseta}
\end{figure}

The spatial constraints act at both local and global level of the
network structure by introducing a distance cost in the establishment
of connections. It is therefore important to look at the effect of
space in global topological quantities such as the betweenness
centrality.  The betweenness centrality of a vertex is determined by
its ability to provide a path between separated regions of the
network.  Hubs are natural crossroads for paths and it is natural to
observe a marked correlation between $g$ and $k$ as expressed in the
general relation $g(k)\sim k^{\mu}$. The exponent $\mu$ depends on the
characteristics of the network and we expect this relation to be
altered when spatial constraints become important. In the present
model, Fig.~\ref{fig:muvseta} clearly shows that this correlation in
fact increases when spatial constraints become large (i.e. when $\eta$
decreases).  This can be understood simply by the fact that the
probability to establish far-reaching short-cuts decreases
exponentially in Eq.~(\ref{sdrive}) and only the large traffic of hubs
can compensate this decay. Far-away geographical regions can thus only
be linked by edges connected to large degree vertices, which implies a
more central role for these hubs.
\begin{figure}
\vspace*{.1cm} 
\centerline{
\includegraphics*[width=0.20\textwidth]{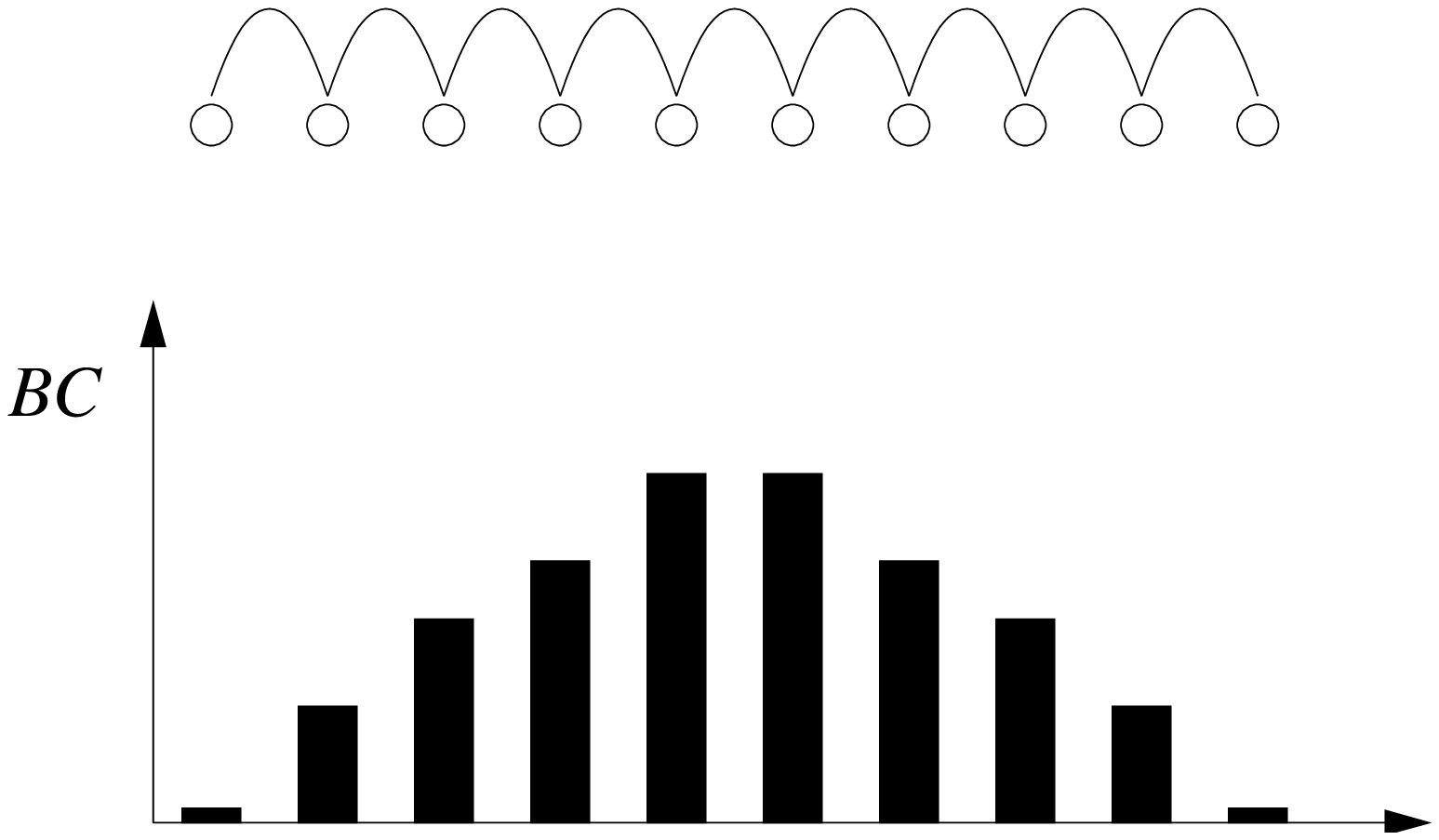}
\includegraphics*[width=0.20\textwidth]{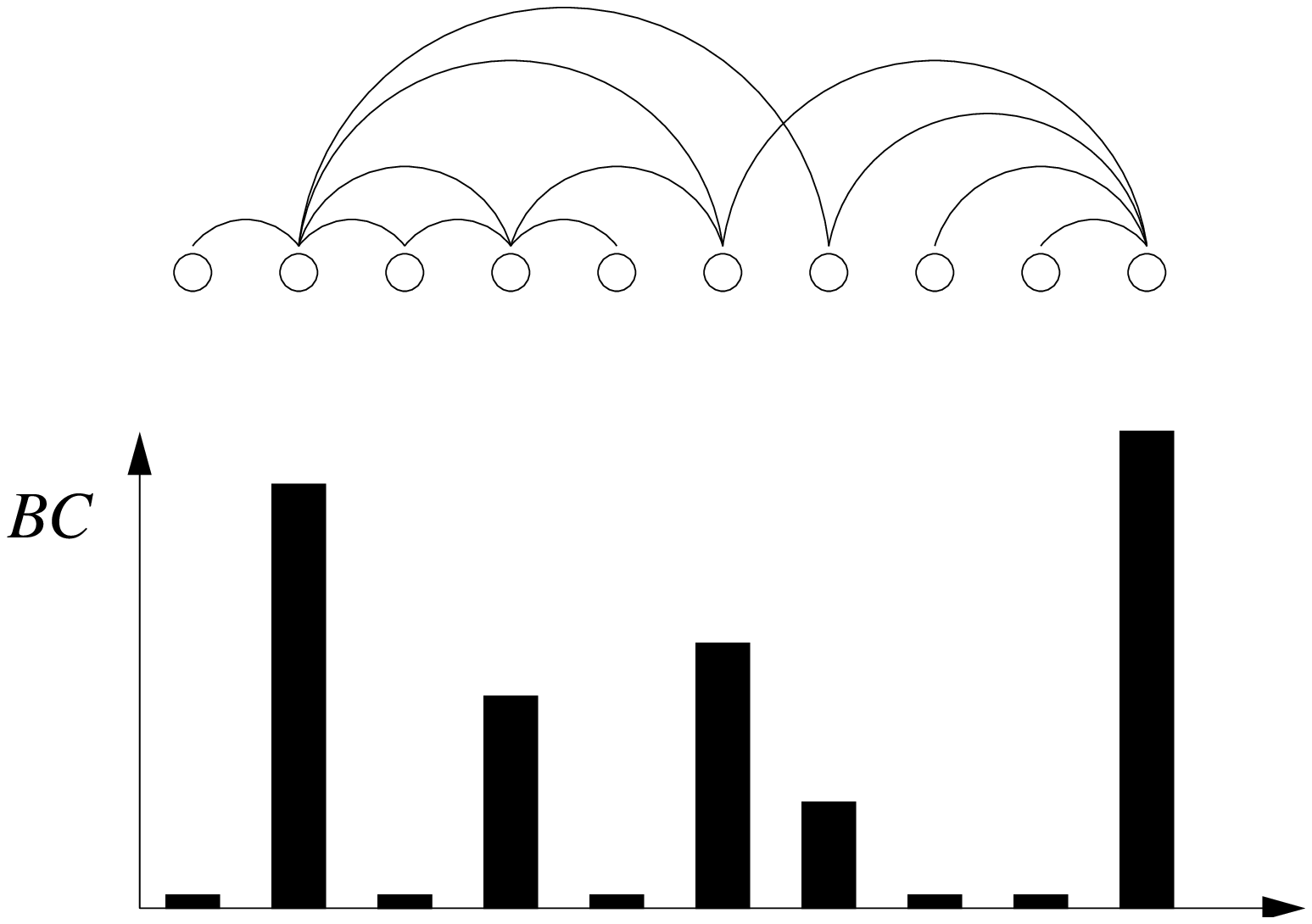}
}
\caption{ (a) Betweenness centrality for the (one-dimensional) lattice
case. The central nodes are close to the barycenter. (b) For a
general graph, the central nodes are usually the ones with large
degree.}
\label{fig:spacebc}
\end{figure}

In order to better understand the effect of space on the properties
of betweenness centrality, we have to explicitly consider the geometry
of the network  along with the topology. In particular, we need to
consider the role of the spatial position by introducing the spatial
barycenter of the network. Indeed, in the presence of a spatial
structure, the centrality of nodes is correlated with their position with
respect to the barycenter $G$, whose location is given
by ${\bf x}_G=\sum_i {\bf x}_i/N$. For a
spatially ordered network---the simplest case being a lattice embedded in a
one-dimensional space---the shortest path
between two nodes is simply the Euclidean geodesic. In a limited
region, for two points lying 
far away, the probability that the shortest path passes
near the barycenter of all nodes is very large. In other words, this implies
that the barycenter (and its neighbors) will have a large
centrality. In a purely topological network with no underlying geography, 
this consideration does
not apply anymore and the full randomness and the disordered small
world structure are completely uncorrelated with the spatial position.
It is worth remarking that the present argument applies in the
absence of periodic boundary conditions that would destroy the
geometrical ordering. This point is illustrated in Fig.~\ref{fig:spacebc} 
in the simple case of a one-dimensional lattice.

\begin{figure}
\vspace*{.1cm} 
\centerline{
\includegraphics*[width=0.30\textwidth]{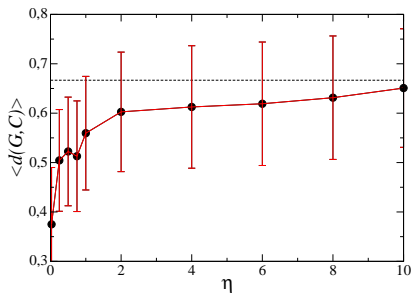}}
\caption{ Average Euclidean distance between the barycenter $G$ of all nodes 
  and the $10$ most central nodes ($C$) versus the parameter $\eta$ (Here
  $\delta=0$, $N=5,000$ and the results are averaged over $50$
  configurations). When space is important (ie.  small $\eta$), the central
  nodes are closer to the gravity center. For large $\eta$, space is
  irrelevant and the average distance tends to the value corresponding to a
  uniform distribution $\langle r \rangle_{unif}=2/3$ (dotted line).  }
\label{fig:rvseta}
\end{figure}

The present model defines an intermediate situation in that we have a
random network with space constraints that introduces a local structure
since short distance connections are favored. Shortcuts and long
distance hops are present along with a spatial local structure that 
clusters spatially neighboring vertices. In Fig.~\ref{fig:rvseta} we
plot the average distance $d(G,C)$ between the barycenter $G$ and the $10$
most central nodes. As expected, as spatial constraints become more important,
the most central nodes get closer to the spatial barycenter of the network.

\begin{figure}
\vspace*{.1cm} 
\centerline{
\includegraphics*[width=0.30\textwidth]{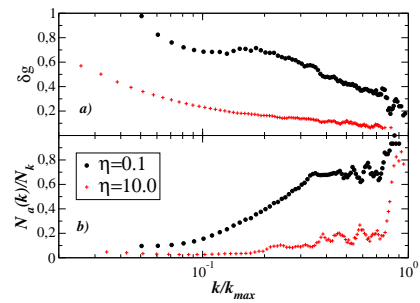}}
\caption{(a) Relative fluctuations of the betweenness centrality 
  versus $k/k_{max}$ for 
  two values of $\eta$ ($N=5,000$ and the results are averaged over
  $50$ configurations and binned). The fluctuations increase when
  $\eta$ decreases (i.e. when spatial constraints increase).
  (b) Number of
  anomalies $N_a(k)$ rescaled by the number of nodes $N_k$ versus
  $k/k_{max}$ for different values of $\eta$. The relative number of
  anomalies is larger when spatial constraints are large,
  especially for large $k$. }
\label{fig:dgvsk}
\end{figure}

Another effect observed when the spatial constraints become important
are the large fluctuations of the BC. Fig.~\ref{fig:dgvsk}a displays
the relative fluctuation
\begin{equation}
\delta g(k)= \frac {\sqrt{\langle \delta g^2(k) \rangle}}
{\langle g(k)\rangle} \, ,
\end{equation}
where $\langle \delta g^2(k)\rangle $ is the variance of the BC and $\langle
g(k)\rangle$ its average (computed for each value of $k$). The value of $\eta$
modifies the degree cut-off and in order to be able to compare the
results for different values of $\eta$ we rescale the abscissa by its maximum
value $k_{max}$. This plot (Fig.~\ref{fig:dgvsk}a) clearly shows that the BC
relative fluctuations increase as $\eta$ decreases and become quite large.
This means that nodes with small degree may have a relatively large BC
(or the opposite), as observed in the air-transportation network (see
Fig.~\ref{fig:bc_na} and~\cite{amaral:2004}). In order to quantify these
``anomalies'' we compute the fluctuations of the betweenness centrality
$\Delta_{RN}(k)$ for a randomized network with the same degree
distribution than the original network and constructed with the Molloy-Reed
algorithm~\cite{molloyreed}. We consider a node $i$ as being ``anomalous'' if
its betweenness centrality $g(i)$ lies outside the interval
$[\langle g(k) \rangle -\alpha\Delta_{RN}(k),
\langle g(k)\rangle +\alpha\Delta_{RN}(k)]$, where we choose
$\alpha\simeq 1.952$ so that the considered interval would represent $95\%$ of
the nodes in the case of Gaussian distributed centralities around the average.
In Fig.~\ref{fig:dgvsk}b, we show the relative number of anomalies
versus $k/k_{kmax}$ for different values of $\eta$. This plot shows
that the relative number of anomalies $N_a(k)/N_k$ increases when the
degree increases and more interestingly strongly increases when
$\eta$ decreases. Note that since for increasing $k$ the number of
nodes $N_k$ is getting small, the results become more noisy.

The results of Figs.~(\ref{fig:muvseta}-\ref{fig:dgvsk}) can be
summarized as follows. In a purely topological growing network,
centrality is strongly correlated with degree since hubs have a
natural ability to provide connections between otherwise separated
regions or neighborhoods~\cite{barthelemy:2003b}. As spatial
constraints appear and become more important, two factors compete in
determining the most central nodes: (i) on the one hand hubs become even
more important in terms of centrality since only a large traffic can
compensate for the cost of long-range connections which implies that
the correlations between degree and centrality become thus even
stronger; (ii) on the other hand, many paths go through the
neighborhood of the barycenter, reinforcing the centrality of
less-connected nodes that happen to be in the right place; this yields
larger fluctuations of $g$ and a larger number of ``anomalies''.

We finally note that these effects are not qualitatively affected by
the weight structure and we observe the same behavior for $\delta=0$
or $\delta\neq 0$.

\section{Conclusions}
\label{sec:V}

In this paper, we have presented a model of growing weighted networks
introducing the effect of space and geometry in the establishment of
new connections. When spatial constraints appear,
the effects on the network structure can be summarized
as follows:

\begin{itemize}
\item{} (i) {\it Effect of spatial embedding on topology-traffic correlations}\\
Spatial constraints induce strong nonlinear correlations between
topology and traffic. The reason for this behavior is that spatial
constraints favor the formation of regional hubs and reinforces
locally the preferential attachment, leading for a given degree
to a larger strength than the one observed without spatial
constraints. Moreover, long-distance links can connect only to hubs,
which yields a value $\beta_{d}>1$ for small enough $\eta$. The
existence of constraints such as spatial distance selection induces
some strong correlations between topology (degree) and
non-topological quantities such as weights or distances.
\item{} (ii) {\it Effect of space embedding on centrality}\\
Spatial constraints also induce large betweenness centrality
fluctuations. While hubs are usually very central, when space is
important central nodes tend to get closer to the gravity center of
all points. Correlations between spatial position and centrality
compete with the usual correlations between degree and centrality,
leading to the observed large fluctuations of centrality at fixed
degree.
\item{} (iii) {\it Effect of space embedding on clustering and assortativity}\\
Spatial constraints implies that the tendency to connect to hubs is
limited by the need to use small-range links. This explains the almost
flat behavior observed for the assortativity. Connection costs also
favor the formation of cliques between spatially close nodes and thus
increase the clustering coefficient.
\end{itemize}

Including spatial effects in a simple model of weighted networks thus yields
a large variety of behavior and interesting effects. This study sheds some
light on the importance and effect of different ingredients such as spatial
embedding or diversity of interaction weights in the structure of large
complex networks and we believe that this attempt of a network typology could
be useful in the understanding and modeling of real-world networks.

\begin{acknowledgments}
  We thank L.A.N. Amaral, E. Chow, S. Dimitrov, R. Guimer\`a, P. de Los
  Rios, T. Pettermann for interesting discussions at various stages of
  this work. A.B and A.V. are partially funded by the European
  Commission - Fet Open project COSIN IST-2001-33555 and contract
  001907 (DELIS).
\end{acknowledgments}



\end{document}